\documentclass[graybox]{svmult}

\usepackage{mathptmx}       
\usepackage{helvet}         
\usepackage{courier}        
\usepackage{type1cm}        
\usepackage{makeidx}         
\usepackage{graphicx}        
\usepackage{multicol}        
\usepackage[bottom]{footmisc}

\makeindex             

\usepackage{amsmath,amssymb,curves,epic}

\begin{document}

\title*{Black Hole Entropy\\and the Problem of Universality}
\author{S.\ Carlip}
\institute{S.\ Carlip \at Department of Physics, 1 Shields Ave., University of California at Davis, 
Davis, CA 95616, USA, 
\email{carlip@physics.ucdavis.edu}}

\maketitle

\abstract*{}

\abstract{To derive black hole thermodynamics in any quantum theory 
of gravity, one must introduce constraints that ensure that a black
hole is actually present.  For a  large class of black holes, the 
imposition of such ``horizon constraints'' allows the use of conformal 
field theory methods to compute the density of states, reproducing the 
correct Bekenstein-Hawking entropy in a nearly model-independent manner.  
This approach may explain the ``universality'' of black hole entropy, the 
fact that many inequivalent descriptions of quantum states all seem to 
give the same thermodynamic predictions.  It also suggests an elegant 
picture of the relevant degrees of freedom, as Goldstone-boson-like 
excitations arising from symmetry breaking by a conformal anomaly 
induced by the horizon constraints.}

\section{The Problem of Universality}
\label{sec:Car1}

Nearly thirty-five years have passed since Bekenstein \cite{Bekenstein}
and Hawking \cite{Hawking} first showed us that black holes are 
thermodynamic objects, with characteristic temperatures
\begin{equation}
T_{\scriptstyle\mathit{H}} = \frac{\hbar\kappa}{2\pi c}
\label{ca1}
\end{equation}
and entropies
\begin{equation}
S_{\scriptstyle\mathit{BH}} = \frac{A}{4\hbar G} \ .
\label{ca2}
\end{equation}
From the start, it was clear that a statistical mechanical description of
these states would be rather peculiar: in contrast to the entropy of an 
ordinary thermodynamic system, black hole entropy is not extensive,
depending on area rather than volume.  Moreover, by Wheeler's famous 
dictum, ``a black hole has no hair'': a classical black hole is determined 
completely by a few macroscopic characteristics, with no apparent room 
for additional microscopic states.  Nevertheless, from the earliest days 
of black hole thermodynamics, the search for a microscopic understanding
has been a vigorous area of research.

Until fairly recently, that search was largely unsuccessful.  Some interesting
ideas were suggested---entanglement entropy of quantum fields across the 
horizon \cite{Bombelli}, or the entropy of quantum fields near the horizon
\cite{tHooft}---but these remained speculative.
Today, in contrast, a great many physicists can tell you, often in great 
detail, exactly what microscopic degrees of freedom underlie black hole
thermodynamics.  The new problem is that they will offer you many
\emph{different} explanations.  Depending on who you ask, black hole 
entropy may count
\begin{itemize} 
\item Weakly coupled string and D-brane states \cite{StromVafa,Peet}
\item Horizonless ``fuzzball'' geometries \cite{Mathur}
\item States in a dual conformal field theory ``at infinity'' \cite{AGMOO,Skenderis}
\item Spin network states crossing the horizon \cite{Ashtekar}
\item Spin network states inside the horizon \cite{Livine}
\item Horizon states in a spin foam \cite{Garcia}
\item ``Heavy'' degrees of freedom in induced gravity \cite{Fursaev}
\item Entanglement entropy  \cite{Bombelli} (maybe ``holographic'' 
  \cite{Ryu,Fursaeva})
\item No local states---it's inherently global \cite{Hawkingb}
\item Nothing---it comes from quantum field theory in a fixed non-quantum
  background \cite{Hawking}, which knows nothing of quantum gravity
\item Maybe something else (points in a causal set in the horizon's domain of 
dependence \cite{Rideout}? 
Kolmogorov-Sinai entropy of strings spreading at the horizon \cite{Ropotenko}?)
\end{itemize}

There is, of course, nothing wrong with a healthy competition among candidates
for the proper description of the quantum black hole.  The relevant degrees of freedom 
are, after all, presumably quantum gravitational---the Bekenstein-Hawking entropy
(\ref{ca2}) involves both $\hbar$ and $G$---and we do not yet have an
established quantum theory of gravity.  But the fact that so many descriptions
give exactly the same answer is a true puzzle.

To see this puzzle more clearly, consider one of the most successful approaches
to black hole entropy, that of weakly coupled string theory.  To count black hole 
microstates a la Strominger and Vafa \cite{StromVafa}, one should proceed as 
follows:
\begin{enumerate}
\item Start with an extremal, supersymmetric, charged black hole;
\item find the horizon area and express it as a function of the charges;
\item ``tune down'' the gravitational coupling to form a weakly coupled 
  string/brane system;
\item count the states in this weakly coupled system, and express their number
  in terms of the charges;
\item argue that supersymmetry (or other properties \cite{Goldstein})
  guarantees that the number of states is the same at strong and weak coupling;
\item compare the results of steps 2 and 4 to determine the entropy as a function
  of the horizon area.
\end{enumerate}

The method is very effective, even away from extremality, and allows the
computation not only of black hole entropy, but of Hawking radiation and even
gray-body factors.  But the fundamental relationship between entropy and area
arises only indirectly, by way of the computation of charges, and this computation
is different for each new type of black hole.  One cannot use the results from, say,
a three-charge black hole in five dimensions to conclude anything about a
four-charge black hole in six dimensions, but must recalculate the entropy and
horizon area for each new case.  Weakly coupled string theory gives the 
Bekenstein-Hawking entropy, but it gives it one black hole at a time.

\section{Conformal Field Theory and the Cardy Formula}
\label{sec:Car2}

The natural question, then, is whether some property of the classical black hole
can explain this ``universality'' by  determining the number of quantum states, 
independent of the details of their description.  This is a lot to ask, and I know of 
only one case in which  such a phenomenon occurs.   Let us therefore take 
a brief detour to explore two-dimensional conformal field theory.

A conformal field theory is a field theory that is invariant under both 
diffeomorphisms  (``general covariance'') and Weyl transformations 
(``local scale invariance'' or ``conformal invariance'') \cite{CFT}.  In two 
dimensions,  one can always choose complex coordinates; such a theory is then 
characterized by two symmetry generators $L[\xi]$ and ${\bar L}[{\bar\xi}]$, 
which generate holomorphic and antiholomorphic diffeomorphisms.  The 
Poisson bracket algebra of these generators is given by the unique central 
extension of the algebra of two-dimensional diffeomorphisms, the Virasoro algebra:
\begin{eqnarray}
\left\{L[\xi],L[\eta]\right\} &=& L[\eta\xi' - \xi\eta']
  + \frac{c}{48\pi}\int dz
  \left( \eta'\xi^{\prime\prime} - \xi'\eta^{\prime\prime}\right)\nonumber \\
\left\{{\bar L}[{\bar\xi}],{\bar L}[{\bar\eta}]\right\} 
  &=& {\bar L}[{\bar\eta}{\bar\xi}' - {\bar\xi}{\bar\eta}']
  + \frac{{\bar c}}{48\pi}\int d{\bar z}
  \left( {\bar\eta}'{\bar\xi}^{\prime\prime} 
  - {\bar\xi}'{\bar\eta}^{\prime\prime}\right) \label{ca3}\\
\left\{L[\xi],{\bar L}[{\bar\eta}]\right\} &=& 0 \ ,\nonumber
\end{eqnarray}
where the central charges $c$ and $\bar c$ (the ``conformal anomalies'') 
depend on the particular theory.  The zero-mode generators $L_0 =
L[\xi_0]$ and ${\bar L}_0 = {\bar L}[{\bar\xi}_0]$ are conserved
charges, roughly analogous to energies; their eigenvalues are commonly
referred to as ``conformal weights'' or ``conformal dimensions.''

In 1986, Cardy discovered a remarkable property of such theories  
\cite{Cardy,Cardyb}.   Given \emph{any} unitary two-dimensional conformal 
field theory for which the lowest eigenvalues $\Delta_0$ of $L_0$ and 
${\bar\Delta}_0$ of ${\bar L}_0$ are nonnegative, the asymptotic density 
of states at large eigenvalues $\Delta$ and $\bar\Delta$ takes the
form
\begin{equation}
\ln\rho(\Delta,{\bar\Delta}) \sim 
  2\pi\sqrt{\frac{(c-24\Delta_0)\Delta}{6}} 
  + 2\pi\sqrt{\frac{({\bar c}-24{\bar\Delta}_0){\bar\Delta}}{6}} \ ,
\label{ca4}
\end{equation}
with higher order corrections that are also determined by the symmetry
\cite{Carlipl,Farey,Birminghama}.  The entropy is thus fixed by 
symmetry, independent of any details of the states being counted.
Note that upon quantization, after making the usual substitutions 
$\{\bullet,\bullet\}\rightarrow[\bullet,\bullet]/i\hbar$ and $L_m \rightarrow L_m/\hbar$,
a classical central charge $c^{\hbox{\scriptsize\it cl}}$ contributes  
$c^{\hbox{\scriptsize\it cl}}/\hbar$ to the quantum central charge, and 
a classical conformal ``charge'' $\Delta^{\hbox{\scriptsize\it cl}}$
contributes $\Delta^{\hbox{\scriptsize\it cl}}/\hbar$ to the quantum 
conformal weight.  The classical piece of a conformal field theory thus
yields a term of order $1/\hbar$ in the entropy (\ref{ca4}), reproducing 
the behavior of the Bekenstein-Hawking entropy (\ref{ca2}).

At first sight, these results seem irrelevant to our problem.  Black holes are 
not typically two dimensional, and neither are they conformally invariant.  
There is a sense, though, in which black holes are \emph{nearly} two dimensional 
and \emph{nearly} conformally invariant near their horizons.  Consider, for
example, a scalar field $\varphi$ near a black hole horizon.   If we write the
metric in ``tortoise'' coordinates
\begin{equation}
ds^2 = N^2(dt^2-dr_*{}^2) + ds_\perp{}^2 \quad \hbox{with
$N\rightarrow0$ at the horizon},
\label{ca5}
\end{equation}
the Klein-Gordon operator becomes
\begin{equation}
 (\Box - m^2)\varphi =
     \frac{1}{N^2}(\partial_t^2 - \partial_{r_*}^2)\varphi + {\cal O}(1) \ ,
\label{ca6}
\end{equation}
and it is evident that the mass and transverse excitations become negligible
as $N\rightarrow0$.  The field is thus effectively described by a two-dimensional 
conformal field theory \cite{Birmingham}.  A similar phenomenon occurs for other 
types of matter, and also, in a sense, for gravity: a generic black hole metric admits
an approximate conformal Killing vector near the horizon \cite{Medved}.

Such an effective two-dimensional description has proven very useful in 
black hole thermodynamics.   Building on old results of Chistensen and Fulling
\cite{Christensen},  Wilczek, Robinson, Iso, Morita, Umetsu, and  
others have recently shown that the Hawking radiation flux \cite{Wilczek} 
and, indeed, the full thermal spectrum \cite{Iso,Isob} can be extracted from
a two-dimensional conformal description, using methods that rely on the
conformal anomalies $(c,{\bar c})$.  Like most derivations of Hawking
radiation, these arguments are based on quantum field theory in a fixed
black hole background.  But as Claudio showed long ago \cite{Teitelboimb},
essentially any effective two-dimensional description of gravity  
also involves a Virasoro algebra, typically with a nonvanishing central
charge.   We might therefore hope that the conformal description could also 
tell us about the statistical mechanics of the black hole states themselves.

\section{2+1 Dimensions}
\label{sec:Car3}

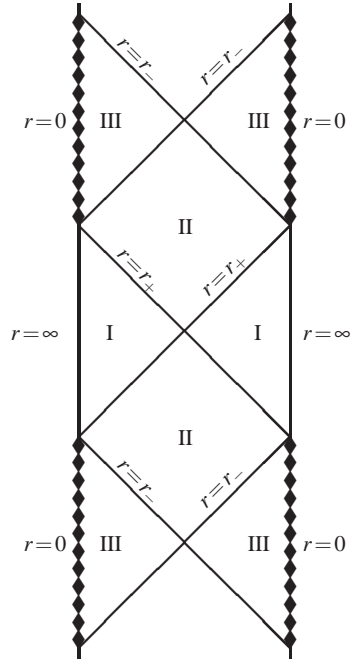
\begin{figure}[t]
\sidecaption[t]
\begin{picture}(210,245)(-75,20)
\setlength{\unitlength}{.8pt}
\thicklines
%
%
\put(0,15){\line(0,1){10}}
\put(100,15){\line(0,1){10}}
\put(0,315){\line(0,1){10}}
\put(100,315){\line(0,1){10}}
\put(0,116){\line(0,1){108}}
\put(100,116){\line(0,1){108}}
%
\dottedline[$\blacklozenge$]{8.5}(0,24)(0,116)
\dottedline[$\blacklozenge$]{8.5}(100,24)(100,116)
\dottedline[$\blacklozenge$]{8.5}(0,224)(0,316)
\dottedline[$\blacklozenge$]{8.5}(100,224)(100,316)
\put(0,20){\line(1,1){100}}
\put(0,120){\line(1,-1){100}}
\put(0,120){\line(1,1){100}}
\put(0,220){\line(1,-1){100}}
\put(0,220){\line(1,1){100}}
\put(0,320){\line(1,-1){100}}
\put(10,66){III}
\put(80,66){III}
\put(47,116){II}
\put(13,166){I}
\put(83,166){I}
\put(47,216){II}
\put(10,266){III}
\put(80,266){III}
\put(-26,66){$r\!=\!0$}
\put(106,66){$r\!=\!0$}
\put(-32,166){$r\!=\!\infty$}
\put(106,166){$r\!=\!\infty$}
\put(-26,266){$r\!=\!0$}
\put(106,266){$r\!=\!0$}
\put(57,86){\rotatebox{45}{$r\!=\!r_-$}}
\put(58,186){\rotatebox{45}{$r\!=\!r_+$}}
\put(57,286){\rotatebox{45}{$r\!=\!r_-$}}
\put(16,106){\rotatebox{-45}{$r\!=\!r_-$}}
\put(16,206){\rotatebox{-45}{$r\!=\!r_+$}}
\put(16,306){\rotatebox{-45}{$r\!=\!r_-$}}
\end{picture}
\caption{The Carter-Penrose diagram for a nonextremal BTZ black hole
 \label{fig1}}
\end{figure}

There is one case in which a conformal field theory derivation of black hole
entropy has been completely successful \cite{Strom,BSS}.  The Ba{\~n}ados-%
Teitelboim-Zanelli black hole \cite{BTZ,Carlip} is a solution of the vacuum 
Einstein equations in three spacetime dimensions with a negative cosmological 
constant $\Lambda=-1/\ell^2$.  Like all vacuum spacetimes in 2+1 dimensions, 
the BTZ geometry has constant curvature, and can be in fact expressed as a 
quotient of anti-de Sitter space by a discrete group of isometries.  Nevertheless, 
it is a real black hole:
\begin{itemize}
\item It has a genuine event horizon at $r=r_+$ and, if the angular momentum is
nonzero, an inner Cauchy horizon at $r=r_-$, where $r_\pm$ are determined
by the mass and angular momentum;
\item it occurs as the end point of the gravitational collapse of matter;
\item its Carter-Penrose diagram, figure \ref{fig1}, is essentially the same as
 that of an ordinary Kerr-AdS black hole;  
\item it exhibits standard black hole thermodynamics, with a temperature and 
entropy given by (\ref{ca1}) and (\ref{ca2}), where the horizon ``area'' is the 
circumference $A=2\pi r_+$.
\end{itemize}

The conformal boundary of a (2+1)-dimensional asymptotically anti-de Sitter 
spacetime is a two-dimensional cylinder, so it is perhaps not surprising that 
the algebra of asymptotic symmetries of the BTZ black hole is a Virasoro 
algebra (\ref{ca3}).  It is rather more surprising that this algebra has a 
central extension, but as Brown and Henneaux showed in 1986 \cite{Brown}, the 
classical central charge, computed from the standard ADM constraint algebra, 
is nonzero:
\begin{equation}
c = \frac{3\ell}{2G} \ .
\label{ca7}
\end{equation}
The appearance of this central charge can be traced back to the need for
boundary terms in the canonical generators of diffeomorphisms, a phenomenon
that we understand largely because of the pioneering work of Claudio and
his collaborators \cite{Regge}.
Moreover, the classical conformal weights $\Delta$ and ${\bar\Delta}$ can be 
calculated in ordinary canonical general relativity, employing the same methods
that are used to determine the ADM mass \cite{Brown}.  Indeed, for the BTZ
black hole, the zero modes of the diffeomorphisms are linear combinations 
of time translations and rotations, and the corresponding conserved quantities 
are linear combinations of the ordinary ADM mass and angular momentum.  
A straightforward calculation gives
\begin{equation}
\Delta = \frac{1}{16G\ell} (r_+ + r_-)^2, \quad 
{\bar\Delta} = \frac{1}{16G\ell} (r_+ - r_-)^2 \ ,
\label{ca8}
\end{equation}
and the Cardy formula (\ref{ca4}) then yields an entropy
\begin{equation}
S = \log\rho\sim \frac{2\pi}{8G}(r_+ + r_-) + \frac{2\pi}{8G}(r_+ - r_-) 
  = \frac{2\pi r_+}{4G} \ ,
\label{ca9}
\end{equation}
which may be recognized as precisely the Bekenstein-Hawking entropy.  

This derivation is one of the first examples of Maldacena's celebrated 
AdS/CFT correspondence \cite{AGMOO}: the entropy of an asymptotically 
anti-de Sitter spacetime is determined by the properties of a boundary conformal 
field theory.  It is also a deeply mysterious result.  Quantum gravity in three spacetime 
dimensions has no local degrees of freedom \cite{Carlipb}, and it is not at all clear 
where one can find enough degrees of freedom to account for the entropy (\ref{ca9}).  
A review of current proposals can be found in \cite{Carlipc}; I will return to 
this question, in a  more general context, in section \ref{sec:Car6}.

The BTZ black hole demonstrates in principle that conformal field theory can be 
used to compute black hole entropy.  Unfortunately, the generalization to higher
dimensions is difficult.  The derivation of \cite{Strom,BSS} depends crucially
on the fact that the conformal boundary of (2+1)-dimensional asymptotically AdS
space is a two-dimensional cylinder, which provides a setting for a two-dimensional
conformal field theory.  No higher-dimensional analog of the Cardy formula is
known,\footnote{Conformal field theory is qualitatively different in two and more
than two dimensions: for $d>2$, the symmetry group has a finite set of generators,
but for $d=2$ it has infinitely many \cite{CFT}.}   so one cannot, at least for now,
use symmetries of a higher-dimensional boundary to constrain the density of
states.  

Moreover, the BTZ computations depend on a symmetry at infinity rather than at 
the horizon.  In 2+1 dimensions this may not matter, since there are no propagating 
degrees of freedom between the black hole and the conformal boundary, but in 
higher dimensions, it is less clear how to isolate black hole degrees of freedom.  
One might argue that a single black hole configuration should make the dominant 
contribution at infinity, but even this is now known to not always be true \cite{Moore}.

Despite these limitations, the BTZ results have proven surprisingly versatile.  In
particular, many near-extremal black holes---including most of the black holes
whose entropy can be computed using weakly coupled string theory---have a 
near-horizon geometry of the form $\mathit{BTZ}\times\mathit{trivial}$,
allowing the application of the BTZ method in a more general setting \cite{Skenderis}.
For generic, nonextremal black holes, though, a more general extension is needed.

\section{Horizons and Constraints}
\label{sec:Car4}

While the conformal analysis of the BTZ black hole does not extend directly to
higher dimensions, it does suggest some interesting directions.  We should, perhaps,
look for a hidden conformal symmetry, of the type discussed in section \ref{sec:Car2},
with a classical central charge; but we should look near the horizon.

To do so, we must first confront a fundamental conceptual issue.  How, in a quantum 
theory of gravity, do we specify that a black hole is present?  In  a semiclassical approach,
this is easy: we fix a background black hole metric and look at quantum fields and
metric fluctuations in that background.  In a full quantum theory of gravity, though,
we cannot do that: the metric is a quantum operator whose components do
not commute, and cannot be simultaneously specified.  We must therefore look for a
more limited set of constraints that are sufficient to guarantee the presence of the
desired black hole while remaining quantum mechanically consistent.  The simplest
way to do this is to add conditions that ensure the presence of a horizon of some
sort---say, an isolated horizon \cite{Ashtekarb}---and study quantum gravity in 
the presence of these additional constraints.  Physically, this amounts to asking 
questions about conditional probabilities: for instance, ``What is the probability of 
detecting a Hawking radiation photon of energy $E$, \emph{given} the presence 
of a horizon of area $A$?''

There are several ways to add such ``horizon constraints,'' which are reviewed in
\cite{Carlipd}.  One approach is to treat the horizon as a sort of boundary.  At
first sight, this seems a peculiar thing to do: a black hole horizon is certainly not
a physical boundary for a freely falling observer.  But a horizon \emph{is} a
hypersurface at  which we can impose ``boundary conditions''--- namely, the 
conditions that it is, in fact, a horizon.  As in the BTZ case, such restrictions 
require boundary terms in the generators of diffeomorphisms, whose presence 
affects their algebra.   It can then be shown that in \emph{any} spacetime of
dimension greater than two, the subgroup of diffeomorphisms in the $r$--$t$ plane
becomes a Virasoro algebra with the right central charges and conformal weights 
to yield the Bekenstein-Hawking entropy \cite{Carlipe,Carlipf,Cvitan}.   

Unfortunately, the diffeomorphisms whose algebra leads to this result are generated 
by vector fields that blow up at the horizon \cite{Dreyer,Koga}, and this divergence 
is poorly understood.   Moreover, this method does not seem to work for the interesting
case of the two-dimensional dilaton black hole.  One can therefore look at a slightly
different approach, in which the ``horizon constraints'' are literally imposed as
constraints in canonical general relativity \cite{Carlipg,Carliph}.

The basic steps of this approach can be summarized as follows:
\begin{enumerate}
\item Dimensionally reduce to the ``$r$--$t$ plane,'' which, as argued in section 
\ref{sec:Car2}, is the relevant setting for near-horizon conformal symmetry.  
Such a reduction is possible even in the absence of spherical or cylindrical symmetry,
although it comes at the expense of an infinite-dimensional Kaluza-Klein gauge group
\cite{Yoon}.  The action then becomes
\begin{equation}
I = \frac{1}{2}\int d^2x\sqrt{g}\left[ \varphi R 
  + V[\varphi] - \frac{1}{2} W[\varphi]h_{IJ}F^I_{ab}F^{Jab}\right]  \ ,
\label{ca10}
\end{equation}
where the dilaton $\varphi$ is the dimensionally reduced remnant of the transverse
area and $F^I$ is the field strength for the usual Kaluza-Klein gauge field $A^I$.
The potentials $V$ and $W$ depend on the details of the higher-dimensional theory,
and need not be further specified.
\item Continue to ``Euclidean'' signature, as Claudio has often advocated 
\cite{Teitelboima}.  The metric can then be written in the form
\begin{equation}
ds^2 = N^2f^2dr^2 + f^2(dt + \alpha dr)^2 \ .
\label{ca11}
\end{equation}
For a black hole spacetime, the horizon shrinks to a point and time becomes
an angular coordinate (see figure \ref{fig2}), with a period $\beta$ determined by the
geometry.\footnote{Claudio and his collaborators were among the first to study such 
black holes in two-dimensional dilaton gravity \cite{Teitelboimc}.}
\begin{figure}[t]
\sidecaption[t]
\linethickness{.3mm}
\begin{picture}(180,170)(40,10)
\setlength{\unitlength}{.85pt}
\put(100,100){\bigcircle{200}}
\put(100,100){\circle*{3}}
\put(100,100){\circle{20}}
\thinlines
\put(140,100){\vector(1,0){60}}
\put(167,101){\small $r$}
\put(90,100){\arc(115,9){27}}
\put(192,154){\vector(-1,2){5}}
\put(200,142){\small $t$}
\thinlines
\put(91,93){\oval(18,18)[br]}
\put(100,93){\vector(0,1){4}}
\put(60,81){\small horizon}
\put(86,109){\oval(18,18)[bl]}
\put(86,100){\vector(1,0){4}}
\put(55,111){\shortstack{\small stretched\\ \small horizon}} 
\end{picture}
\caption{A ``Euclidean'' black hole spacetime \label{fig2}}
\end{figure}
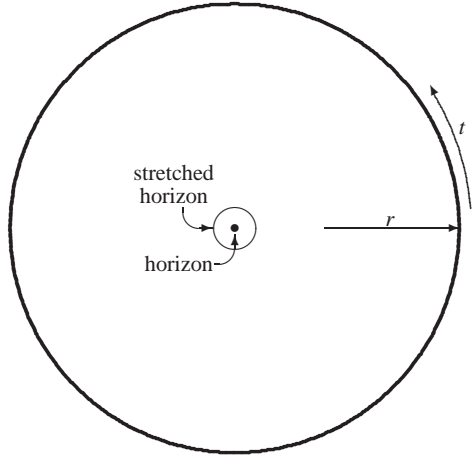
Rather than evolving in $t$, we borrow a trick from conformal field theory 
\cite{CFT} and evolve radially, starting at a ``stretched horizon'' just outside
$r=0$.
\item Find the ordinary ADM-style constraints, which take the form
\begin{align}
&{\cal H}_\parallel = {\dot\varphi}\pi_\varphi - f{\dot\pi}_f = 0\nonumber\\
&{\cal H}_\perp = f\pi_f\pi_\varphi + f\left(\frac{\dot\varphi}{f}\right)^\cdot
  -\frac{1}{2}f^2{\hat V} = 0 
\qquad\hbox{with}\ \ {\hat V} = V + \frac{h^{IJ}\pi_I\pi_J}{W}\nonumber\\
&{\cal H}_I = {\dot\pi}_I - c^J{}_{IK}A^K\pi_J = 0 \ . \label{ca12}
\end{align}
These can be combined to form Virasoro generators 
\begin{align}
L[\xi] &= \frac{1}{2}\int dt\xi({\cal H}_\parallel + i{\cal H}_\perp) \nonumber\\
{\bar L}[{\bar\xi}] &= \frac{1}{2}\int dt{\bar\xi}({\cal H}_\parallel - i{\cal H}_\perp) \  ,
\label{ca13}
\end{align}
which satisfy the algebra (\ref{ca3}) with vanishing central charge.
\item Determine the geometrical quantities that characterize the black hole:
\begin{alignat}{2}
&\mathrm{the\ expansion} &\qquad  s &= \varphi\vartheta 
  = f\pi_f - i{\dot\varphi} \label{ca14}\\
&\mathrm{the\ surface\ gravity} & {\hat\kappa}  
&= \pi_\varphi - i{\dot f}/{f} + f^2\frac{d\omega}{d\varphi} \ .\nonumber
\end{alignat}
The surface gravity is not unique---in standard general relativity, it depends on
the normalization of the Killing vector at the horizon \cite{Ashtekarb}, which  
here appears as a conformal factor $\omega$ that will be determined later.
\item Impose horizon constraints to ensure that our initial surface is a stretched
horizon.  As Claudio noted in \cite{Teitelboima}, the actual horizon is determined by 
the conditions $s={\bar s}=0$.  A stretched horizon with surface gravity 
${\hat\kappa}_H$ is naturally specified by the slightly loosened conditions
\begin{align}
K &= s - a({\hat\kappa} - {\hat\kappa}_H) = 0 \nonumber\\
{\bar K} &= {\bar s} - a({\bar{\hat\kappa}} - {\bar{\hat\kappa}}_H) = 0 \ ,
\label{ca15}
\end{align}
where the constant $a$ will be determined below.
\item Note that the horizon constraints $K$ and $\bar K$ do not commute with 
the Virasoro generators (\ref{ca13}), which are therefore not symmetries of the 
constrained system.  Cure this problem by using the Bergmann-Komar formulation 
of Dirac brackets \cite{Bergmann}.  Let $\{K_i\}$ be a set of constraints  for which 
the inverse $\Delta_{ij}$ of $\{K_i,K_j\}$ exists (in Dirac's language, a set of 
second class constraints).  Then for any observable $\cal O$, the new observable
\begin{equation}
{\cal O}^* = {\cal O} - \sum_{i,j}\int dudv\{{\cal O},K_i(u)\}\Delta_{ij}(u,v)K_j(v)
\label{ca16}
\end{equation}
will have vanishing Poisson brackets with the $K_i$.  Since ${\cal O}^*$ differs 
from $\cal O$ only by a multiple of the constraints $K_i$, the two are physically 
equivalent.  The Poisson bracket $\{{\cal O}_1^*,{\cal O}_2^*\}$ can be shown 
to be equal to the Dirac bracket of ${\cal O}_1$ and ${\cal O}_2$.
\item Work out the Poisson algebra of the modified Virasoro generators $L^*[\xi]$
and ${\bar L}^*[{\bar\xi}]$.  The conformal factor $\omega$ in (\ref{ca14}) and the 
constant $a$ in (\ref{ca15}) are both fixed by the requirement that these brackets 
be ``nice,'' and in particular that they reduce to an ordinary Virasoro algebra at 
the horizon.  Choosing modes 
\begin{equation}
\xi_n = \frac{\beta}{2\pi}e^{2\pi int/\beta}
\label{ca17}
\end{equation}
for the vector fields used to smear the Virasoro generators, we obtain central
charges and conformal weights
\begin{equation}
c = {\bar c} = \frac{3\varphi_H}{4G} ,\qquad 
\Delta = {\bar\Delta} 
= \frac{\varphi_H}{16G}\left(\frac{\kappa_H\beta}{2\pi}\right)^2 \ .
\label{ca18}
\end{equation}
\item Use the Cardy formula (\ref{ca4}) to obtain an entropy
\begin{equation}
S = \frac{2\pi\varphi_H}{4G}\left(\frac{\kappa_H\beta}{2\pi}\right)  \ .
\label{ca19}
\end{equation}
This is \emph{almost} the correct Bekenstein-Hawking entropy; it differs from
the correct expression by a factor of $2\pi$.  I believe this factor has a simple
physical explanation: entropy should count the black hole degrees of freedom at 
a fixed time, but because of our choice of radial evolution, we have computed the 
entropy at the horizon for \emph{all} times, effectively integrating over a circle of 
circumference $2\pi$.
\end{enumerate}

\section{Universality Again}
\label{sec:Car5}

Now, however, let us recall our original motivation, which was to understand the
``universality'' of black hole entropy.  If the conformal field theory/horizon
constraint picture is to explain this universality, it must be the case that the
symmetry of the preceding section is secretly present in all of the other
derivations of black hole entropy.  This is certainly not obvious, but there are 
a few hopeful signs.

Let us first compare the horizon constraint method to the conformal approach to
the BTZ black hole described in section \ref{sec:Car3}.  We can start by comparing
the central charges and conformal weights:
\begin{alignat}{3}
&& &\qquad\mathbf{\underline{BTZ}} 
       & &\qquad\quad\mathbf{\underline{Horizon\ CFT}} \nonumber \\
&{\mathit{{modes}}}\qquad 
      & &\qquad\xi_n \sim e^{in(t\pm\ell\phi)/\ell}\qquad & &\quad\qquad\xi_n 
      \sim e^{in\kappa_H t}\nonumber\\
&c & &\qquad\ \frac{3\ell}{2G} 
      & &\qquad\ \ \ \  2\pi\cdot\frac{3\varphi_H}{4\pi G} \label{ca20} \\
&\Delta,\ {\bar\Delta} & &\qquad\ \frac{(r_+\pm r_-)^2}{16G\ell} 
     & &\qquad\!\!\! 2\pi\cdot\frac{\varphi_H}{32\pi G}
     \left(\frac{\kappa_H\beta}{2\pi}\right)^2 \nonumber
\end{alignat}
While the entropies agree, it appears that the central charges and conformal weights
do not.  In fact, though, these disagreements can be traced to two simple sources
\cite{Carliph}: the periodicities of the modes do not match, and the BTZ results 
are based on a coordinate system that is not corotating at the horizon, as one
would desire for dimensional reduction.  Once these differences are accounted for,
the central charges and conformal weights agree precisely.  As noted in section
\ref{sec:Car3}, the BTZ approach applies also to most of the black holes that can
be exactly analyzed with weakly coupled string theory, so this agreement is a 
significant step.

For loop quantum gravity, the connection is less clear.  There is, however, an
interesting coincidence that may point toward something deeper.  The horizon
states of a spin network described in \cite{Ashtekar} are characterized by a constrained 
$\mathrm{SL}(2,{\mathbb R})$ Chern-Simons theory with coupling constant 
$k=iA/8\pi\gamma G$, where $\gamma$ is the Immirzi parameter.  Any 
three-dimensional Chern-Simons theory has an associated two-dimensional 
conformal field theories, a Wess-Zumino-Witten model that appears, for example, 
in the description of boundary states \cite{Witten}.  In the present case,
this conformal field theory is Liouville theory,  and its central charge is approximately 
$6k$.  If we choose Ashtekar's original self-dual formulation of loop quantum gravity 
\cite{Ashtekarc}, for which $\gamma=i$, this central charge agrees precisely
with the value obtained by the horizon constraint approach.  

The central charge (\ref{ca18}) also matches that of the ``horizon as boundary'' 
approach of \cite{Carlipf}, and the conformal weights can be obtained as a Komar 
integral, as suggested in a slightly different context by Emparan and Mateos 
\cite{EmpMat}.  I believe it should also be possible to relate this method to the 
Euclidean path integral approach to black hole entropy; work on this question is 
in progress.

\section{What are the States?}
\label{sec:Car6}

If near-horizon conformal symmetry really provides a universal explanation for
black hole statistical mechanics, it had better \emph{not} give us a unique
description of the relevant microstates.  The problem, after all, is that many
different microscopic descriptions seem to yield the same macroscopic thermal
properties; picking out one ``right'' description would miss the point.
Nevertheless, it is possible that the derivation of section \ref{sec:Car4} might
give a useful \emph{effective} description of the microscopic degrees of freedom.

Consider the standard Dirac treatment of constraints in quantum mechanics. 
A set of classical (first class) constraints $L[\xi] = {\bar L}[{\bar\xi}] =0$ translates
 to a quantum restriction on the space of states:
\begin{equation}
L[\xi]|\mathrm{phys}\rangle = {\bar L}[{\bar\xi}]|\mathrm{phys}\rangle = 0 \ .
\label{ca21}
\end{equation}
But in the presence of a central charge, such a restriction is inconsistent with
the Virasoro algebra (\ref{ca3}).  This is not new, of course, and it is well known
how to fix the problem \cite{CFT}; for example, one can require that only the positive
frequency pieces of $L[\xi]$ and ${\bar L}[{\bar\xi}]$ annihilate physical states.  The
net result, though, is that some states that would have been unphysical
in the absence of a central charge must now be considered physical.  Equivalently
\cite{Carlipi}, the presence of boundaries or constraints can remove gauge
degeneracies among otherwise physically equivalent states, turning ``would-be
gauge transformations'' into new dynamical degrees of freedom.

This phenomenon may have first been observed by Claudio.  In an underappreciated
passage in \cite{Teitelboimb}, he points out that the presence of a central charge
in dilaton gravity is quantum mechanically consistent, but results in the appearance
of a new degree of freedom.  In the present context, we are imposing the constraints 
(\ref{ca15}) only at the horizon, so it is only there that a central charge appears,
but the new horizon degree of freedom is essentially the same as Claudio's.

As Kaloper and Terning have observed \cite{Kaloper}, this process is also somewhat
reminiscent of the Goldstone mechanism, in which a spontaneously broken symmetry
gives rise to massless excitations in the direction of the ``broken'' generators.
Here, of course, the broken symmetry is a gauge symmetry, and the corresponding
degrees of freedom are therefore new.  But as in the Goldstone mechanism, the
pattern of symmetry breaking may give us a universal effective description of
the degrees of freedom, while not touching on their ``real'' structure in terms 
of the fundamental underlying quantum gravitational states.

For asymptotically anti-de Sitter spacetimes in three dimensions, an explicit 
description of the symmetry breaking and the corresponding degrees of freedom 
at infinity is possible \cite{Carlipj}.  The resulting effective field theory is a
Liouville theory.  This two-dimensional conformal field theory has the correct 
central charge and conformal weights to yield the Bekenstein-Hawking entropy
via the Cardy formula, but there is still a debate as to whether it really contains 
enough degrees of freedom \cite{Carlip}.  A similar induced action can be found
in five-dimensional asymptotically anti-de Sitter gravity \cite{Aros}, although
the problem of counting states has not been solved.  One might hope for a more 
general result in arbitrary dimension, perhaps focusing on the horizon rather
than infinity; Claudio is responsible for an interesting effort in this direction 
\cite{Teitelboima}.

One avenue for further research may be to look more carefully at the path 
integral measure, which is in some sense a count of the number of states, in the 
presence of a Virasoro algebra with a nonzero central charge.  It is known
that when second class constraints $\{C_i\}$ are present, the path integral 
acquires a Fadeev-Popov-like determinant $\det|\{C_i,C_j\}|^{1/2}$ \cite{HennTeit}.
For a Virasoro algebra, this is
\begin{equation}
D = \det\left| \{L_m,L_n\}\right|^{1/2} \ .
\label{ca22}
\end{equation}
A naive evaluation of this expression, using the algebra (\ref{ca3}), \emph{almost}
gives the Cardy formula: one finds $D\sim\exp\{2\pi\sqrt{6\Delta/c}\}$, which
differs from (\ref{ca4}) by a flip from $c/6$ to $6/c$.  This is a bit too simple, though,
since the Virasoro algebra contains an $\mathrm{SL}(2,{\mathbb R})$ subgroup,
generated by $\{L_0,L_{\pm1}\}$, whose algebra remains first class.  This adds a 
large degeneracy, increasing the density of states; we really need to evaluate a 
determinant of the form
\begin{equation}
D= 
\det \left|-\frac{c}{12}\frac{d^3\ }{dx^3} 
    + \frac{d\ }{dx}L + L\frac{d\ }{dx}\right|^{1/2}
\quad \hbox{with}\ \ L = L_0 + L_1e^{2ix} + L_{-1}e^{-2ix} 
\label{ca23}
\end{equation}
and trace over appropriate  $\mathrm{SL}(2,{\mathbb R})$ states.
It is not yet clear  whether this approach is still too naive; it may be that we need 
more detailed information about the $\mathrm{SL}(2,{\mathbb R})$ 
representations than can be obtained from a constraint analysis alone.

\section{What Next?}
\label{sec:Car7}

While I have given some evidence for the proposal that black hole thermodynamics
is effectively determined by near-horizon conformal symmetry, the hypothesis
remains very far from being proven.  I can think of two main directions to
proceed.

First, we should try to connect the near-horizon symmetry more closely to other
derivations of black hole entropy.  In loop quantum gravity, does the numerical 
coincidence discussed in section \ref{sec:Car5} have any deeper significance?
Is there a way of using the associated Liouville theory to count states?  In the 
``fuzzball'' approach to black holes in string theory \cite{Mathur}, no single 
configuration is expected to have a near-horizon conformal symmetry (or, indeed, 
a horizon); can a sum over configurations recover such a symmetry?  In induced 
gravity \cite{Fursaev}, a connection to conformal field theory is already
known \cite{Frolov}; can it be tied to the near-horizon symmetry discussed 
here?  Can the determinant (\ref{ca23}) be evaluated, and will it give the correct 
density of states?  Does the spin foam method of \cite{Garcia}, which relies on 
a treatment of the horizon as an effective boundary, contain a hidden conformal 
symmetry?

Second, we should keep in mind that there is more to black hole thermodynamics
than the Bekenstein-Hawking entropy.  As I noted in section \ref{sec:Car2},
the intensity and spectrum of Hawking radiation can be obtained from an
effective two-dimensional theory near the horizon, using conformal field
theory methods applied to matter rather than gravity \cite{Wilczek,Iso,Isob}.
It is natural to hope that these matter degrees of freedom can be coupled to
the near-horizon gravitational degrees of freedom to obtain a dynamical
description of Hawking radiation.  In 2+1 dimensions, Emparan and Sachs
have shown that something of this sort may be possible \cite{Emparan}: a 
classical scalar field can be coupled to the conformal boundary degrees of 
freedom of the BTZ black hole, and conformal methods then yield the correct
description of Hawking radiation.  If this result could be generalized to arbitrary
dimensions, with a coupling at the horizon, it would provide very strong
evidence for the conformal description of black hole thermodynamics.

\begin{acknowledgement}
This work was supported in part by Department of Energy grant
DE-FG02-91ER40674.
\end{acknowledgement}

\end{document}